# Suppression of nucleation density in twisted graphene domains grown on graphene/SiC template by sequential thermal process

*Yao Yao[1*], Taiki Inoue[1], Makoto Takamura[2], Yoshitaka Taniyasu[2] and Yoshihiro Kobayashi[1*]*

[1] Osaka University, 2-1 Yamadaoka, Suita, Osaka 565-0871, Japan,

[2] NTT Basic Research Laboratories, 3-1 Morinosato-Wakamiya, Atugi-shi, Kanagawa 243-0198, Japan

ABSTRACT

We investigated the growth of twisted graphene on graphene/silicon carbide (SiC-G) templates by metal-free chemical vapor deposition (CVD) through a sequential thermal (ST) process, which exploits the ultraclean surface of SiC-G without exposing the surface to air before CVD. By conducting control experiments with SiC-G templates exposed to air (AirE process), structural analysis by atomic force microscopy revealed that the nucleation density of CVD graphene (CVD-G) was significantly suppressed in the ST process under the same growth condition. The nucleation behavior on SiC-G surfaces is observed to be very sensitive to carbon source concentration and process temperature. The nucleation on the ultraclean surface of SiC-G prepared by the ST process requires higher partial pressure of carbon source compared with that on the surface by the AirE process. Moreover, analysis of CVD-G growth over a wide temperature range indicates that nucleation phenomena change dramatically with a threshold temperature of 1300°C, possibly due to arising of etching effects. The successful synthesis of twisted few-layer graphene (tFLG) was affirmed by Raman spectroscopy, in which analysis of the G' band proves a high ratio of twisted structure in CVD-G. These results demonstrate that metal-free CVD utilizing ultraclean templates is an effective approach for the scalable production of large-domain tFLG that is valuable for electronic applications.

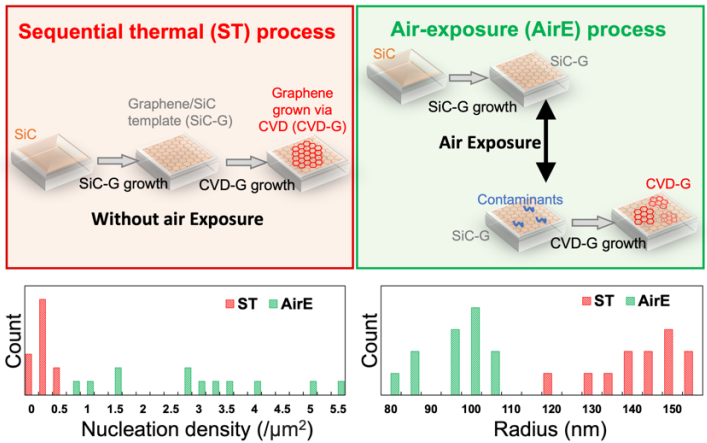

Schematic diagram of SiC-G and CVD-G growth of graphene in ST process and AirE process. Statistical distribution of nucleation density and radius of graphene domains in both ST process and AirE process.

1. Introduction

Twisted few-layer graphene (tFLG) exhibits massless fermion behavior and linear dispersion of the electronic band near the Fermi level due to the misorientation of Bernal-stacked or rhombohedral-stacked graphene [1-2]. This behavior extends the electronic versatility of such two-dimensional (2D) carbon materials. tFLG with diverse angles can result in many unusual properties, such as van Hove singularities [3], topological flat bands [4], quantum Hall effects [5-6], and enhanced superlattice potentials [7]. Owing to the weak interlayer coupling [4] and screening effect [8], graphene is less influenced by the charged impurities on a substrate [9]. Thus, tFLG is a promising candidate to be applied to field-effect transistors, optoelectronics devices, and microsupercapacitors, as it enhances the transport characteristics [10-13].

A common method for fabricating tFLG is the multiple transfer of monolayer graphene prepared by mechanical exfoliation [14], but the exfoliated graphene transferred by polymer stamps is unscalable. Chemical vapor deposition (CVD), as a method for growing graphene on a large area, has been reported in numerous studies showing that CVD growth on metal foil as a catalytic surface can produce scalable graphene [15-16]. Unfortunately, fabricating more than two layers of twisted graphene on copper foil is challenging [17], and multilayer AB-stacked graphene on nickel foil is commonly grown [16]. Thus, non-catalyst layer-by-layer growth using monolayer graphene as a template can overcome both the limitations of the layer number and the twisted structure at the same time [18-19].

We demonstrated the twisted structure of CVD-grown graphene (CVD-G) layers on monolayer graphene templates by scanning transmission electron microscopy and G' peak analysis of Raman spectroscopy [11,20]. Moreover, using graphene from the thermal decomposition of silicon carbide (SiC-G) as a template, CVD-G grown on SiC-G with a twisted bilayer structure was observed with a scanning tunneling microscope [21]. The method of growing graphene on a graphene template lays the foundation for the study of tFLG applicable to high-performance electrical devices. However, the scattering effect of charged impurities in wedding cake-shaped multilayer graphene reduces carrier mobility, limiting the high quality of tFLG field-effect transistors [20]. Hence, studying the morphology of tFLG and understanding the layer-by-layer growth mechanism are the keys to improving the quality of graphene.

The process to achieve layer-by-layer growth is influenced by several factors, encompassing reaction conditions such as temperature, duration, and carrier gas flow rate [22-23] in addition to aspects like interface mismatches, surface defects, and substrate surface cleanliness [24-26]. A previous study of grown tFLG revealed that the nucleation density is too high to allow sufficient space for the lateral growth of the graphene domain, which refers to graphene formed by binding carbon adatoms extending along the edges from a single nucleation site [21]. With the locally high nucleation probability, graphene is grown into multilayer domain islands due to successive nucleation in the new layer [23]. Therefore, reducing the nucleation probability is crucial for the layer-by-layer growth of CVD-G. When the two growth processes of SiC-G and CVD-G were performed, the samples were taken out of the growth chambers [21]. Consequently, the air-exposure (AirE) process is considered to lead to the deposition of contaminants on the SiC-G surface, which may be one of the factors affecting the excessive nucleation and island-shaped growth of CVD-G. The effects of the surface cleanliness of template graphene should be investigated in detail, and ideal layer-by-layer growth conditions should be created to grow high-quality tFLG.

In this study, we examined an effective approach to growing tFLG with reduced nucleation density by restricting the exposure of template graphene to ambient air before CVD. We performed growth processes of SiC-G and CVD-G in the same chamber to create a sequential thermal (ST) process. As a result, the nucleation density of CVD-G decreased significantly, and the multilayer domains grew to have a larger radius and lower height compared to the case of the AirE process. This effect was attributed to the absence of air contaminants on the graphene templates during the ST process. Additionally, we discussed the temperature dependence and etching effect based on the suppression of nucleation density. We also proposed a growth model to systematically explain the nucleation and growth mechanism of CVD-G on graphene templates. This approach provides a basis for manufacturing high-quality tFLG consisting of large-domain single-crystal graphene.

2. Method

We conducted two growth stages in preparing tFLG: one of template monolayer SiC-G and one of CVD-G. A schematic diagram of the growth stages is shown in Figure 1. The SiC-G growth stage involves ultra-high-temperature annealing to obtain graphene on SiC as a stable template for the following stage. The CVD-G growth stage involves growing graphene by CVD on the template SiC-G at high temperature while supplying ethanol as a carbon source. As depicted in Figure 1 (a), the process of exposing the sample to air between the two stages [21] is referred to as the AirE process. Conversely, in the ST process, the two growth stages are carried out in the same chamber, which is isolated from the air environment, as illustrated in Figure 1 (b). It is essential to ensure sufficient cooling and a continuous vacuum when transitioning between the two stages of the ST process.

For the SiC-G growth stage, commercial 6H-SiC (0001) substrates (on-axis, semi-insulating purchased from II-VI Advanced Materials) were annealed at 1700 °C and 40 kPa for 10 min in an argon atmosphere [27]. In the CVD-G growth stage, graphene layers were synthesized without a metal catalyst by flowing a mixture of ethanol/argon (typically 1 sccm/100 sccm) at a total pressure of 10.5 kPa, corresponding to the ethanol partial pressure of 105 Pa. To compare the initial growth stage with respect to graphene nucleation of the ST and AirE processes, we conducted growth of CVD-G at 1200 °C for 10 min in both processes. To obtain more details about the difference in graphene nucleation between the two processes, the dependence of ethanol partial pressure was systematically investigated in both the AirE and ST processes. By adjusting the mixture ratio of ethanol to argon, the partial pressure of ethanol was varied from 15 Pa to 530 Pa, while temperature and growth time were kept at 1400 ºC and 10 min, respectively. Furthermore, to investigate the CVD-G growth behavior in the ST process, we investigated a wide range of temperatures from 1000 ºC to 1500 ºC for 25 min at an ethanol partial pressure of 105 Pa. All the high-temperature treatments were conducted using an infrared heating furnace (SR1800G, Thermo Riko) which is a cold-wall system, where the sample stage is heated to high temperatures while the chamber walls are not.

The grown tFLG was characterized with several analytical techniques. The atomic force microscopy (AFM) (AFM5100N, Hitachi High-Tech) observations were conducted in the dynamic force mode under atmospheric conditions at room temperature. A Raman spectrometer (HR800, Horiba) was used to analyze the twisted structure and layer number of the CVD-G, using laser excitation of 532 nm with a 100× objective lens (typical spot diameter: ~1 μm). A previously developed method [28] was employed to subtract the spectra from the SiC substrates. The Raman spectra presented in this study were normalized to the characteristic peaks of SiC substrates. By comparing the G-band of SiC-G and that of samples after CVD-G growth, it was possible to estimate the growth amount and layer number of the CVD-G.

3. Results and discussion

3.1 Graphene nucleation by ST and AirE processes

To examine the impact of surface cleanliness, we performed a set of control experiments comparing the ST process with the AirE process, as shown in Figure 2. These experiments were conducted under optimized conditions at 1200 °C with an ethanol partial pressure of 105 Pa for 10 min, which were chosen to minimize the occurrence of CVD-G domain coalescence for the initial growth stage with respect to graphene nucleation. We investigated the growth morphology including the nucleation density, domain radius, and domain height of CVD-G in both ST and AirE processes. A total of 12 sets of AFM images, each measuring 5 μm × 5 μm, were captured from the ST and AirE processes, respectively, for the statistical analysis. Figures 2 (a)-(c) represent one set of examples in the ST process, and Figures 2 (d)-(f) represent one set of examples in the AirE process.

As shown in Figure 2 (a), (b), (d), (e), the isolated domains and coalesced domains of CVD-G can be distinguished. The isolated domains, which refer to graphene formed from a single nucleation site and with approximately circular shapes, are indicated by arrows. A singular peak observed at the center of the domain is regarded as the nucleation site, as exemplified in Figure 2 (c), which corresponds to the bright spot seen in the magnified phase image illustrated in Figure 2 (b). In the case of the coalesced domains with irregular-shaped structures, the number of the nucleation sites was estimated by counting the bright spots in phase images. The magnified image of Figure 2 (e) shows an example of one coalesced domain with five nucleation sites. The height was evaluated based on the cross-sections (Fig. 2c and f) of both the isolated and coalesced domains. To clearly demonstrate the ability of lateral growth under different surface states in the two processes, the radius was evaluated exclusively by analyzing the cross-sections (Fig. 2c and f) of the isolated domains.

The statistical data (Fig. 2g) show that the ST process yields significantly fewer nucleation sites compared to the AirE process. The CVD-G domains formed in the ST process are fewer and dispersed throughout the surface (Fig. 2a), while the CVD-G domains formed in the AirE process are numerous and locally coalesced (Fig. 2d). Upon examining the cross-sections of the CVD-G domains, the statistical results (Fig. 2h and i) show that CVD-G from the ST process presents a lower height and larger radius, indicating a greater potential for the formation of laterally grown large-domain graphene compared to that from the AirE process.

In addition to the comparison under optimized conditions, we also expanded the comparison between the ST and AirE processes under a range of varying pressure conditions for 10 min at 1400 °C. Figure 3(a)-(h) shows the AFM topography of the two processes. In Figure 3 (a), (e), and (f), no graphene was grown. In (b), monolayer 2D graphene domains were grown. In Figure 3 (c), (d), (g), and (h), multilayer graphene domains were grown, forming multilayer graphene on the surface. The coverage determined from the AFM images (Fig. 3i) obviously exhibits growth within different partial pressure ranges. This observation suggests that the critical saturations of the carbon sources for nucleation differ for the two processes. The nucleation threshold in ethanol partial pressure for the ST process (53-105 Pa, Fig. 1f, g) is observably higher than that for the AirE process (15-53 Pa, Fig. 1a, b). Figure 3 (j) presents the Raman spectra that correspond to the AFM images shown in Figure 1(a)-(h). At 53 Pa, the intensified G-band and G'-band in the AirE process are attributed to the growth of CVD-G, while the spectra of the ST process demonstrate only the monolayer SiC-G template. The Raman result is consistent with the AFM observation. Over 53 Pa, the results of Raman intensity indicating lower nucleation density in the ST process than in the AirE process are also demonstrated.

We hypothesize that the effect of suppressing graphene growth in the ST process is due to the prevention of surface contaminants. In the AirE process, due to the interaction with the SiC (0001) substrates and carbon-rich buffer layers, the surface of SiC-G usually adsorbs contaminants, such as adsorbed oxygen, hydrocarbon, or water [29], when it is exposed to the ambient environment. The contaminants existing on the surface of SiC-G decompose at high temperatures during the CVD-G growth stage. On the surface of SiC-G, the decomposition of contaminants provides active sites with low surface energy [30], or where dangling bonds may exist [31], leading to the locally high nucleation density of CVD-G. Therefore, the extended diffusion length of adsorbed carbon atoms on a pristine surface provides a rational explanation for the decreased nucleation density in the ST process [32].

3.2 Detailed study of ST process

3.2.1 Temperature dependence

To investigate the growth mechanism of CVD-G in the ST process in detail, we performed a parametric study on the growth temperature. Figure 4 (a)-(l) shows the CVD-G growth morphology from AFM images as a function of growth temperature for the ST process. Figure 4 (m) is a representative cross-sectional view of a graphene domain, which corresponds to the dotted line N-N' in Figure 4 (l). By employing the same statistical method described in Section 3.1, Figure 4 (n) illustrates the average radius of the isolated domains, revealing an increase in domain radius with growth temperature. Moreover, as shown in Figure 4 (o), (p), the height and nucleation density were determined from AFM topography images, including the isolated and coalesced domains. The density of CVD-G domains decreases and the height of CVD-G increases as the temperature increases. All trends of height, domain radius, and nucleation density are interrupted at 1300 °C with the disappearance of CVD-G. Furthermore, in the temperature range of 1100-1300 °C, isolated domains show single- or few-layer growth. At temperatures above 1300 °C, isolated domains show larger multilayer growth (Fig. 4n, o), and striking pits are observed at the center of all domains. As indicated in Figure 4 (m), the depth of the pit is approximately 5 nm. This value is significantly larger than the layer distance of the graphene connected to SiC by the buffer layer [33].

Figure 5 shows the Raman spectra of the ST process samples and a sample of only SiC-G growth. The Raman spectrum of the only SiC-G growth sample exhibits characteristic peaks of monolayer graphene corresponding to the G-band around 1600 cm$^{-1}$ and the G'-band around 2734 cm$^{-1}$. These features also align with the characteristics of epitaxial monolayer graphene on SiC substrates [34]. The Raman spectrum of the sample after CVD-G growth at 1000 °C and 1300 °C is almost identical to that of only SiC-G. This confirms the absence of any CVD-G growth at this temperature and is consistent with the statistical results from AFM images. Within the temperature range of 1100-1200 °C, there is an increase in the intensity of the G-band, which indicates the growth of CVD-G. The morphologies of the samples observed by AFM are consistent with the Raman spectroscopy results. As the temperature continues to rise above 1300 °C, the significant enhancement of the G-band serves as evidence of the extensive growth of CVD-G. Moreover, for the area completely covered by graphene (domain size must be larger than the typical spot diameter of the Raman spectrometer: ~1 μm), we analyzed the ratio of twisted CVD-G at 1200 °C (Fig. 5b). The G'-band of the spectrum was fitted with three Lorenz curves (Supporting information) to calculate the proportion of AB-stacking and twisted stacking to estimate the ratio of twisted graphene [35]. As a result, the twisted ratio at 1200 °C was determined to be 75.6%, which is significantly higher than those of few-layer graphene obtained by metal-catalyzed CVD [36-37], where there is a preference for the growth of AB-stacking graphene. The high ratio of the twisted structure agrees with the monolayer graphene growth with a random moiré pattern in our previous study [21]. It should be emphasized that the present study demonstrates multilayer CVD-G growth while maintaining a high twisted ratio, which confirms the effectiveness of our approach for obtaining tFLG in a scalable manner.

3.2.2 Modeling of growth behavior

In 2D nucleation, which occurs at a surface or interface, the chemical potential should decrease with increasing temperature according to the energetics based on thermodynamics [38-39]. The nucleation density typically decreases as the temperature increases [40]. In the cold-wall system for graphene nucleation and growth, the decomposition of ethanol increased with the temperature. Although the surface with higher temperature received more carbon adatoms from ethanol, the acceleration of carbon adatom desorption was more dominant for the overall nucleation process, resulting in a low nucleation probability. Thus, the nucleation density decreased and the domain size increased with increasing temperature until CVD-G disappeared at 1300 °C, as demonstrated in the ST process (Fig. 6a-c).

However, at temperatures higher than 1300 °C, the appearance of CVD-G beyond thermodynamic explanation led us to consider an additional mechanism. Previous studies of graphene growth on copper foils have reported the importance of etching effects [41-42], which induce a partial removal of graphene by the decomposition of ethanol molecules. The carbon atoms of ethanol also compensate for the removal, facilitating the restoration of the original graphene, and graphene nucleation occurs on the restored graphene. In our model, as the temperature rises in the cold-wall system, this decomposition of ethanol becomes more pronounced, subsequently elevating the concentrations of oxygen-containing species. These species play an active role in surface etching effects, inducing more defects in SiC-G. Consequently, there is an escalated requirement for supplementary carbon atoms to counterbalance the etching of SiC-G. Therefore, when the temperature exceeds the threshold, as depicted in T > 1300 °C of Figure 6 (d), the higher number of produced oxygen-containing species partially etch away SiC-G. Since the SiC substrate without a graphene cover sublimates as the temperature increases [43], the sublimation of SiC occurs under the etched area of SiC-G.

In this scenario, we speculated that due to the inability of restoration to match the CVD-G removal of etching effects, when SiC sublimation is initiated, surface pits tend to develop. The decomposition of SiC leaves carbon atoms in the vicinity of the pits, resulting in a higher local concentration of carbon adatoms that act as a carbon source for graphene growth. For the graphene domain shown in Figure 4 (m), we estimated the volume of the pit part and the graphene part based on the AFM morphology image and further calculated the number of carbon atoms corresponding to the volume to be $4.08 \times 10^6$ and $1.57 \times 10^8$, respectively, almost in the same order of magnitude. Therefore, we inferred that carbon atoms sublimating from pit regions participate in the local formation of multilayer CVD-G domains [44].

4. Conclusion

An ST process was proposed to improve the growth of twisted graphene on graphene templates from SiC without a metal catalyst. By changing the degree of ethanol decomposition, which was controlled by adjusting the partial pressure of ethanol or temperature of the substrate, different growth behaviors of graphene were observed in the ST and AirE processes. The growth of large-domain layer-by-layer tFLG relies on nucleation with uniformly low density. In this regard, the ST process with a clean surface is a key step to effectively prevent locally high-density nucleation. Raman spectroscopy verified the high ratio of the twisted structure of graphene grown in the ST process. We presented a model incorporating the etching effects of surface graphene, SiC-G restoration, and CVD-G growth to explain the unexpected growth at higher temperature. This research contributes to the fundamental understanding of the tFLG growth mechanism, and the low nucleation density observed in this study is expected to be advantageous for the development of large-scale tFLG synthesis and functional graphene devices.

FIGURES

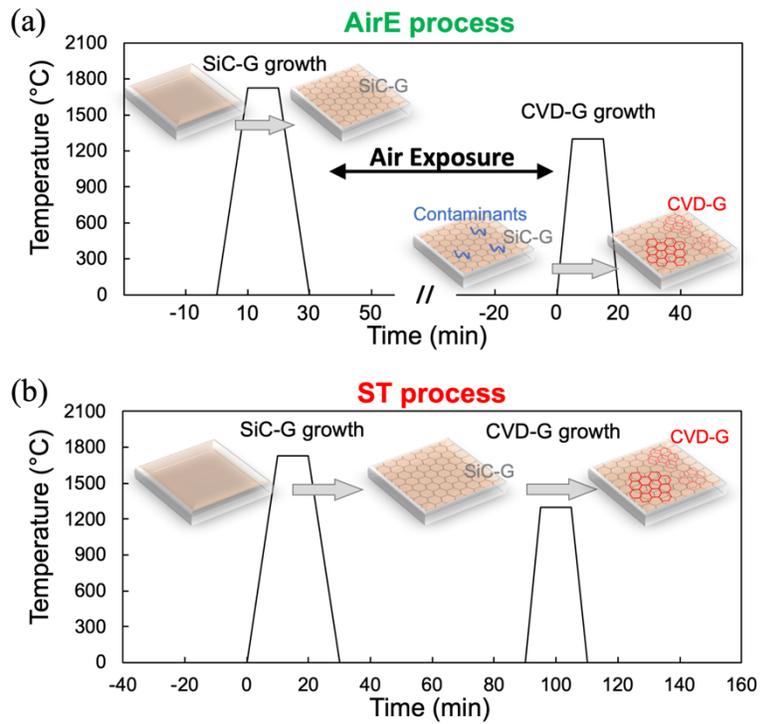

**Figure 1.** Schematic diagram of SiC-G and CVD-G growth of graphene in (a) AirE process and (b) ST process.

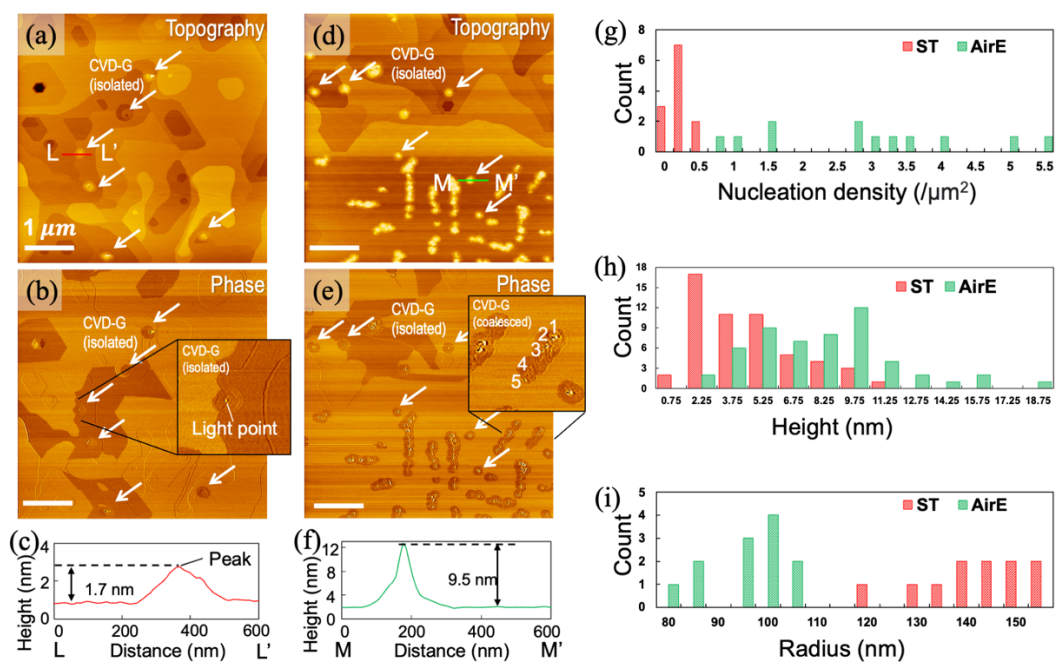

**Figure 2.** AFM images observed from samples after CVD growth at 1200°C for 10 min with ethanol partial pressure of 105 Pa in (a-c) ST process and (d-e) AirE process. (c) Cross-sectional view of red line segment L-L' shown in (a). (f) Cross-sectional view of green line segment M-M' shown in (d). Statistical distribution of CVD-G (g) nucleation density, (h) height of isolated and coalesced domains, and (i) average radius of isolated domains from 12 sets of AFM images in both ST process and AirE process.

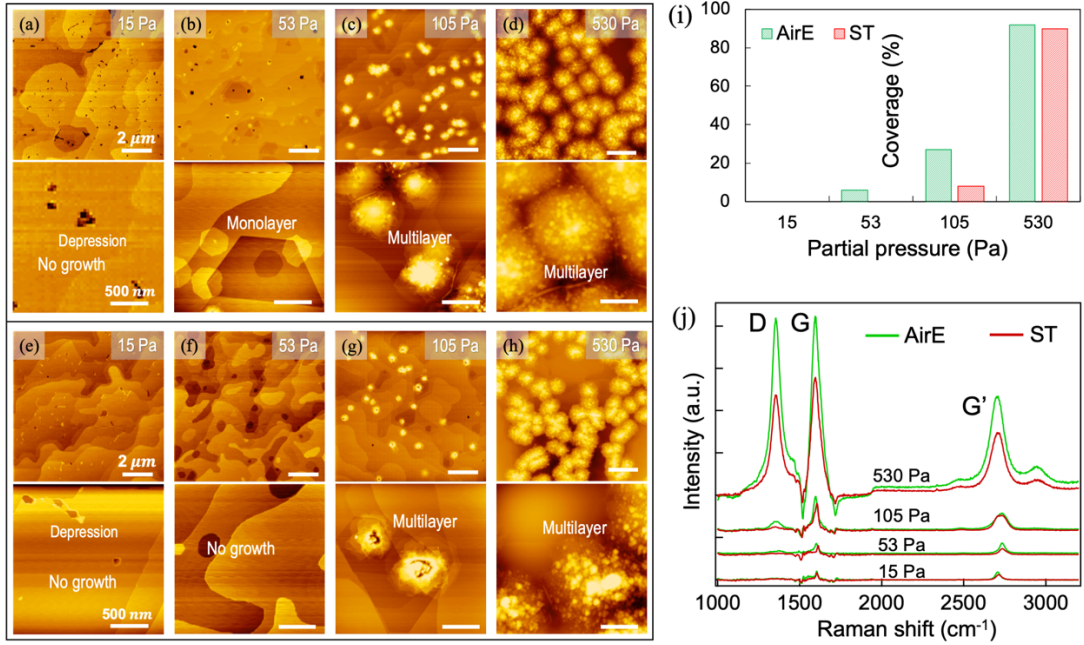

**Figure 3.** AFM images observed from samples after CVD growth at 1400 ºC for 10 min with various partial pressures of ethanol in both (a-d) AirE process and (e-h) ST process. Comparison of AirE and ST processes in (i) coverage of grown CVD-G and (j) Raman spectra.

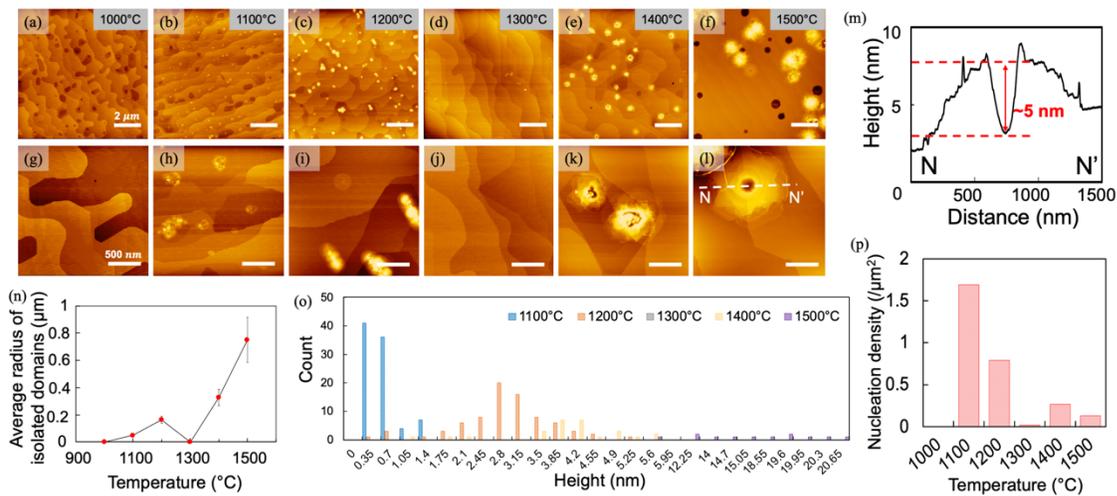

**Figure 4.** AFM images observed from samples after CVD growth in ST process on graphene templates/SiC substrates for 25 min with ethanol partial pressure of 105 Pa at (a) 1000 ºC, (b) 1100 ºC, (c) 1200 ºC, (d) 1300 ºC, (e) 1400 ºC, and (f) 1500 ºC. (g-l) Magnified AFM images of (a-f). (m) Cross-sectional view of dotted line segment N-N' shown in (l). (n) Temperature dependence of average radius of isolated domains observed from grown CVD-G from AFM images. Statistical distribution of (o) height and (p) nucleation density of CVD-G domains evaluated from AFM images at different temperatures.

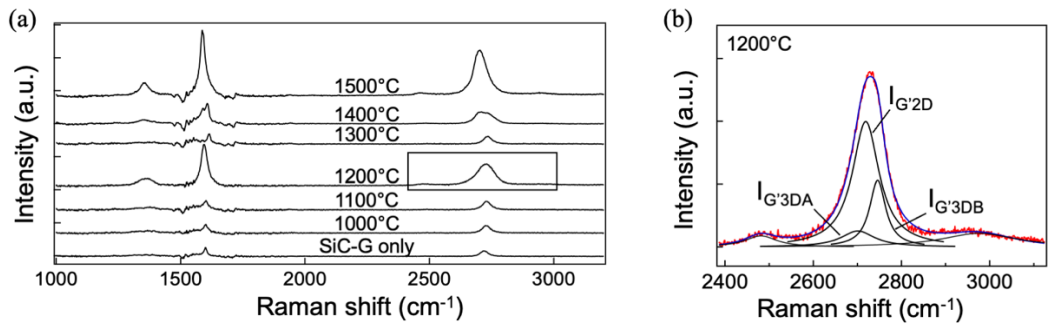

**Figure 5.** (a) Raman spectra of SiC-G grown by ST process at each temperature. (b) Three fitted Lorenz curves of G'-band in Raman spectrum at 1200 ºC. All these spectra are the results after subtraction of SiC substrate peaks and normalization.

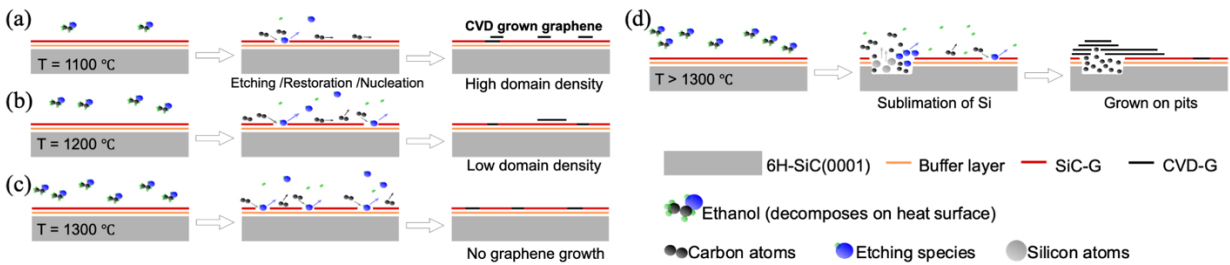

**Figure 6.** Schematic model of CVD-G growth at (a) 1100 ºC, (b) 1200 ºC, and (c) 1300 ºC for ST process based on thermodynamics and reported research [41-42]. Speculation model (d) at temperature greater than 1300 ºC for ST process.

ASSOCIATED CONTENT

**Supporting Information**.

The following files are available free of charge.

Specific information of Raman spectrum analysis of on twisted ratio of CVD-G (PDF)

AUTHOR INFORMATION

**Corresponding Author**

* yaoyao@ap.eng.osaka-u.ac.jp * kobayashi@ap.eng.osaka-u.ac.jp

**Author Contributions**

The manuscript was written through contributions of all authors. All authors have given approval to the final version of the manuscript.

**Funding Sources**

JSPS KAKENHI (JP16K13639, JP15H05867, JP17H05336, JP19H04545, JP17H02745, and JP21H01763) and JST SPRING (JPMJSP2138).

ACKNOWLEDGMENT

This research was supported by JSPS KAKENHI (JP16K13639, JP15H05867, JP17H05336, JP19H04545, JP17H02745, and JP21H01763) and JST SPRING (JPMJSP2138).

ABBREVIATIONS

tFLG, twisted few-layer graphene; SiC-G, graphene grown by thermal decomposition of silicon carbide; CVD, chemical vapor deposition; CVD-G, graphene grown by CVD; ST process, sequential thermal process; AirE process, air-exposure process.

**Supporting information: Evaluation of twisted structure ratio**

By analyzing the shape of the G'-band, it is possible to calculate the proportion of AB stacking to twisted stacking in the stacked structure of graphene [1]. The G'-band of the spectrum are fitted with three Lorenz curves, G'$_{2D}$ for two-dimensional (2D) graphene component, G'$_{3DA}$ and G'$_{3DB}$ for three-dimensional (3D) graphite component, to calculate the AB-stacking proportion:

$$R = \frac{I_{G'\,3DB}}{I_{G'\,3DB} + I_{G'\,2D}} \quad (1)$$

($I_{G'3DA} : I_{G'3DB} = 1:2$)

Then, the twisted ratio is represented by 1−R.

The fitting conditions for this study were set as follows by measuring the G'$_{2D}$ band peak position of monolayer SiC-G and G'$_{3DA}$, G'$_{3DB}$ of graphite, respectively.

$$G'_{3DA} \sim G'_{2D}\ 18\ cm^{-1}$$

$$G'_{3DB} \sim G'_{2D}\ 27\ cm^{-1}$$

($I_{G'3DA} : I_{G'3DB} = 1:2$)

In this study, peak fitting of the G'-band was performed using Igor Pro (manufactured by HULINKS) as analysis software.

Additionally, at a higher ethanol pressure of 530 Pa at 1400 ºC for 10 min, in which multilayer graphene was grown to cover almost the entire surface, the ratios of twisted structure in tFLG are also calculated by fitting the peak shape of the G'-band. Both AirE and ST processes show twisted structures with ratios exceeding 90%.